\documentclass{PoS}

\usepackage{bm}

\title{Bottom hadrons from lattice QCD with domain wall and NRQCD fermions}

\ShortTitle{Bottom hadrons from lattice QCD with domain wall and NRQCD fermions}

\author{\speaker{Stefan Meinel}$\:^a$, William Detmold$^{b,c}$, C.-J. David Lin$^{d,e}$, Matthew Wingate$^a$\\ \\
\llap{$^a$}DAMTP, University of Cambridge, Wilberforce Road, Cambridge CB3 0WA, UK\\
\llap{$^b$}Department of Physics, College of William \&  Mary, Williamsburg,
  VA 23187-8795, USA\\
\llap{$^c$}Jefferson Laboratory, 12000 Jefferson Avenue, Newport News, VA 23606, USA\\
\llap{$^d$}Institute of Physics, National Chiao-Tung University, 
Hsinchu 300, Taiwan\\
\llap{$^e$}Physics Division, National Centre for Theoretical Sciences,
Hsinchu 300, Taiwan\\
\\ \\
E-mail: \email{S.Meinel@damtp.cam.ac.uk}}

\abstract{Dynamical 2+1 flavor lattice QCD is used to calculate the masses of bottom hadrons,
including $B$ mesons, singly and doubly bottom baryons, and for the first time also the triply-bottom
baryon $\Omega_{bbb}$. The domain wall action is used for the up-, down-, and strange quarks
(both valence and sea), while the bottom quark is implemented with non-relativistic QCD. A calculation
of the bottomonium spectrum is also presented.}

\FullConference{The XXVII International Symposium on Lattice Field Theory - LAT2009\\
		 July 26-31 2009\\
		 Peking University, Beijing, China}

\def\sfrac#1#2{{\textstyle\frac{#1}{#2}}}

\def\Dzero{$\rm D0\hspace{-1.15ex}\slash\hspace{0.5ex}$}

\begin{document}

\section{Introduction}

Lattice studies of hadrons containing $b$ quarks are important for several reasons. One major motivation
is flavor physics, where non-perturbative calculations of hadronic matrix elements for electroweak transitions
are required. Secondly, lattice QCD can predict masses of hadrons that have not
yet been observed experimentally. A few singly-bottom baryons have been found so far, and more
results are expected from the LHC. Most recently, the $\Omega_b$ baryon was discovered at Fermilab.
There are now two incompatible results for its mass, obtained by the \Dzero \cite{Abazov:2008qm} and
CDF \cite{Aaltonen:2009ny} collaborations. Lattice QCD can contribute to resolve this discrepancy.

A number of unquenched calculations of bottom baryon masses have been done recently
\cite{Lewis:2008fu,Burch:2008qx,Na:2008hz,Detmold:2008ww,Lin:2009rx} (see also \cite{Aubin:2009yh} for a review presented at this conference).
It is important to perform independent determinations of the same quantities with different lattice formulations in order
to test universality. In this work, the domain wall fermion action (with $L_s=16,\:\:M_5=1.8$) is used for
both the valence and sea $u$-, $d$- and $s$ quarks, while the $b$ quark is treated with non-relativistic QCD (NRQCD).
Compared to the static heavy-quark action, which was used in \cite{Burch:2008qx, Detmold:2008ww,Lin:2009rx},
NRQCD has the advantage that it is not limited to systems containing only a single $b$ quark. Also, spin splittings
which would vanish in the static limit can be calculated.

This work makes use of the $V=24^3\times 64$ gauge configurations generated by the RBC and UKQCD collaborations
\cite{Allton:2008pn}. There are four different ensembles with pion masses ranging from about 672 to 331 MeV;
the lattice spacing is approximately 0.11fm.

The form of the lattice NRQCD action used here is the same as in Ref.~\cite{Gray:2005ur}, where the bottomonium spectrum
was calculated on MILC gauge configurations with AsqTad sea quarks and L\"uscher-Weisz gluons. The RBC/UKQCD ensembles use
different actions for both the sea quarks (domain wall) an the gluons (Iwasaki), and it is therefore a useful test
of universality to compute the bottomonium spectrum again on these lattices before moving on to do heavy-light calculations.
This was done in Ref.~\cite{Meinel:2009rd}. In addition to tests of the lattice actions, this work provided an
accurate tuning of the bare $b$ quark mass and an independent determination of the lattice spacing. The main results
are summarized in Sec.~\ref{sec:bottomonium} below; the reader is referred to \cite{Meinel:2009rd} for the details.

Then, Sec.~\ref{sec:bottom_hadrons} goes on to describe the calculation of the bottom hadron spectrum, including
$B$ mesons, singly- and doubly-bottom baryons, and the triply-bottom $\Omega_{bbb}$. The heavy-light calculations
are still in progress, and here only results for $am_{\rm light}=0.005$, $am_{\rm strange}=0.04$ and limited
statistics are shown. The full, chirally extrapolated results will be presented in a forthcoming publication.

\section{Bottomonium}

\label{sec:bottomonium}

The first step was the tuning of the bare $b$ quark mass. When using NRQCD, all energies obtained from fits
to hadronic two-point functions are shifted by some common constant, as the rest mass is not included in
the theory. Thus, to tune the $b$ quark mass it is convenient to consider
the \emph{kinetic mass}
\begin{equation}
M_{\rm kin} \equiv \frac{\bm{p}^2-\left[E(\bm{p})-E(0)\right]^2}{2\left[E(\bm{p})-E(0)\right]} \label{eq:mkin}
\end{equation}
\begin{table}
\begin{minipage}{.48\linewidth}
\begin{center}
\begin{tabular}{ccccccc}
\hline\hline
$am_b$ &  & $aM_\mathrm{kin}(\eta_b)$ &  &  $\displaystyle\begin{array}{c}
\Upsilon(2S)-\Upsilon(1S) \\ \mathrm{splitting} \end{array}$ \\
\hline
$2.30$  &&  $4.988(12)$  &&  $0.3258(47)$ \\
$2.45$  &&  $5.281(13)$  &&  $0.3242(46)$ \\
$2.60$  &&  $5.575(13)$  &&  $0.3231(54)$ \\
\hline\hline
\end{tabular}
\caption{\label{tab:mb_tuning} $\eta_b(1S)$ kinetic mass and $\Upsilon(2S)-\Upsilon(1S)$ splitting for
three values of the bare $b$ quark mass (lattice units). Errors are statistical/fitting only.}
\end{center}
\end{minipage}
\hfill
\begin{minipage}{.48\linewidth}
\begin{center}
\begin{tabular}{lcccc}
\hline\hline
$am_l$ &   & $a^{-1}_{2S-1S}$ \hspace{1ex} (GeV) \\
\hline
$0.005$  &&  $1.740(25)(19)$    \\
$0.01$   &&  $1.722(38)(19)$    \\
$0.02$   &&  $1.708(92)(19)$    \\
$0.03$   &&  $1.72(12)(2)\phantom{00}$    \\
\hline\hline
\end{tabular}
\caption{\label{tab:lattice_spacing} Results for the inverse lattice spacings of the different ensembles,
obtained from the $\Upsilon(2S)-\Upsilon(1S)$ splitting.}
\end{center}
\end{minipage}
\end{table}
of the hadron. This is based on the relativistic continuum dispersion relation, which is in fact very close
to the lattice dispersion relation in the case considered here: as demonstrated in \cite{Meinel:2009rd},
the \emph{speed of light} is compatible with 1 within statistical errors of less than 0.25\% for lattice
momenta $a\bm{p}=\bm{n}\cdot 2\pi/L$ up to $\bm{n}^2=12$; equivalently $M_\mathrm{kin}$ shows no dependence
on $\bm{p}$ within errors.

\begin{figure}
\begin{minipage}{.48\linewidth}
\centerline{\includegraphics[width=\linewidth]{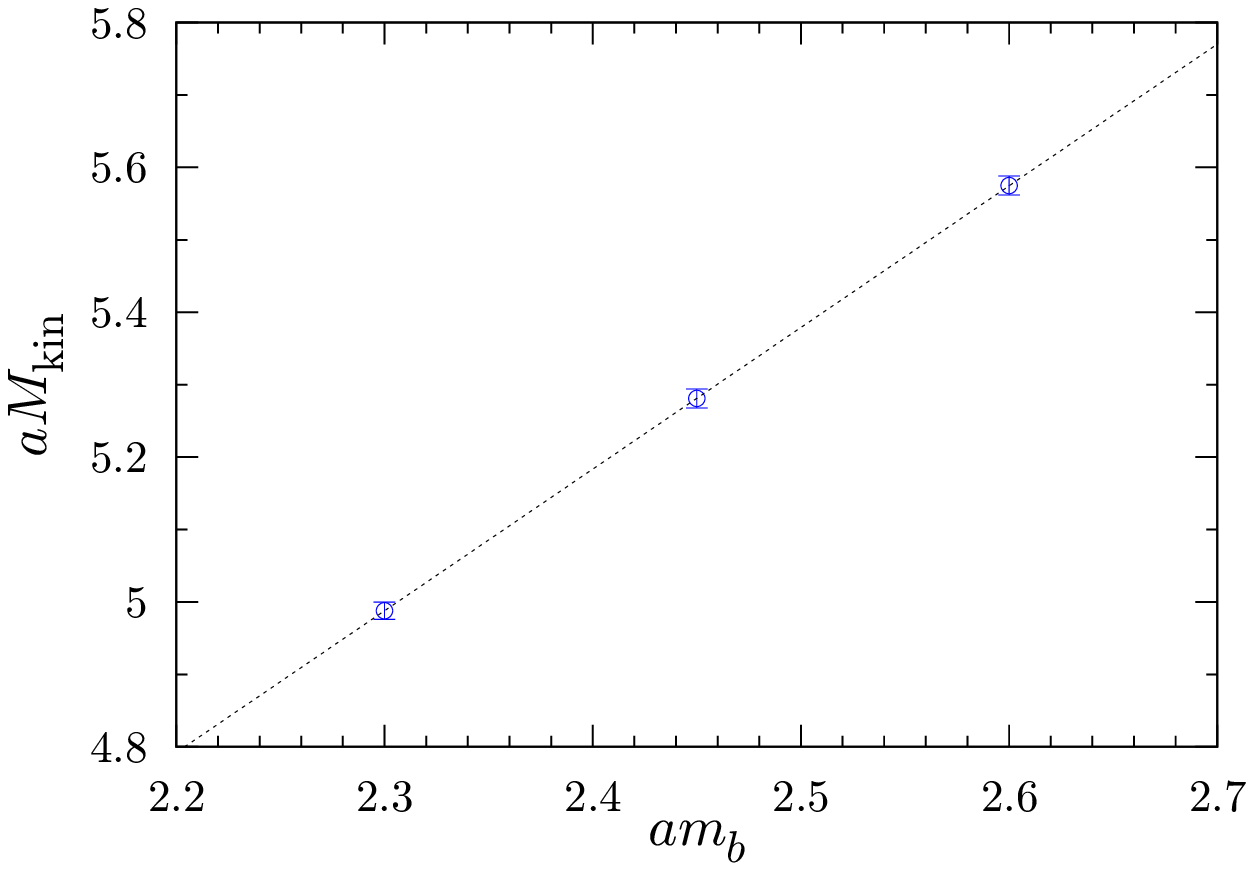}}
\caption{$\eta_b(1S)$ kinetic mass vs $a m_b$. Errors are statistical/fitting only. The line shows a linear fit.}
\label{fig:mkin_vs_mb}
\end{minipage}
\hfill
\begin{minipage}{.48\linewidth}
\centerline{\includegraphics[width=\linewidth]{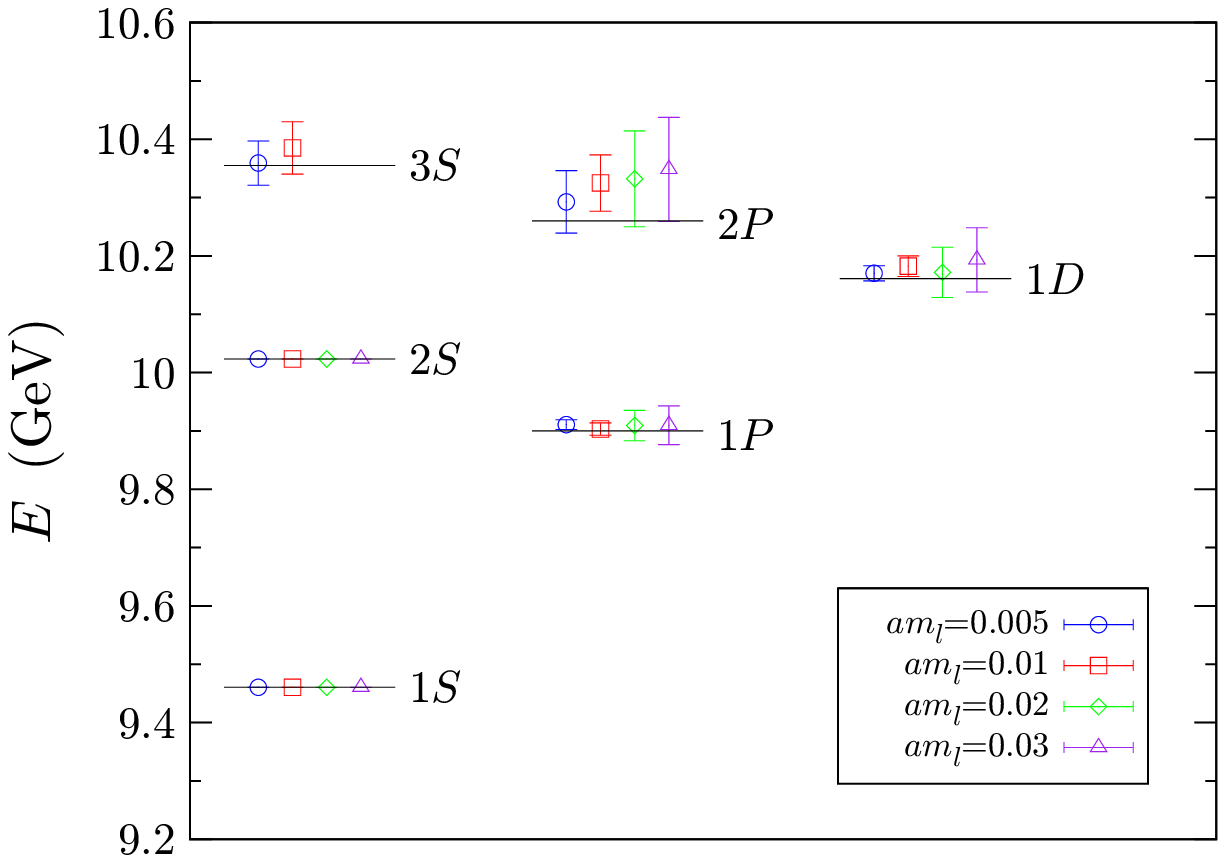}}
\caption{Radial and orbital energy splittings in bottomonium. Errors are statistical/fitting only.}
\label{fig:radial_orbital_vs_experiment}
\end{minipage}
\end{figure}

Table \ref{tab:mb_tuning} shows the kinetic mass of the $\eta_b$ meson for three different values of $am_b$,
on the $am_l=0.005$ ensemble. As can be seen in Fig.~\ref{fig:mkin_vs_mb}, the data are compatible with a
linear dependence in the range considered. Fitting the function $aM_\mathrm{kin}=A+B\cdot am_b$
gives $A=0.489(25)$, $B=1.956(11)$. Also shown in Table~\ref{tab:mb_tuning} is the $\Upsilon(2S)-\Upsilon(1S)$
energy splitting, which is found to be nearly independent of $am_b$. The $\Upsilon(2S)-\Upsilon(1S)$ splitting
is furthermore expected to have very small systematic errors, and is therefore an ideal quantity to set the
lattice scale by comparing to the experimental value of $0.56296(40)$ GeV \cite{Amsler:2008zzb}. Then,
using the experimental value of the $\eta_b$ mass, $9.389(5)$ GeV \cite{Aubert:2008vj},
one can solve for the value of $m_b$ that gives the correct kinetic mass in physical units. This gives
\begin{equation}
a m_b = 2.514(36).  \label{eq:amb}
\end{equation}

\begin{figure}
\begin{minipage}{.48\linewidth}
\centerline{\includegraphics[width=\linewidth]{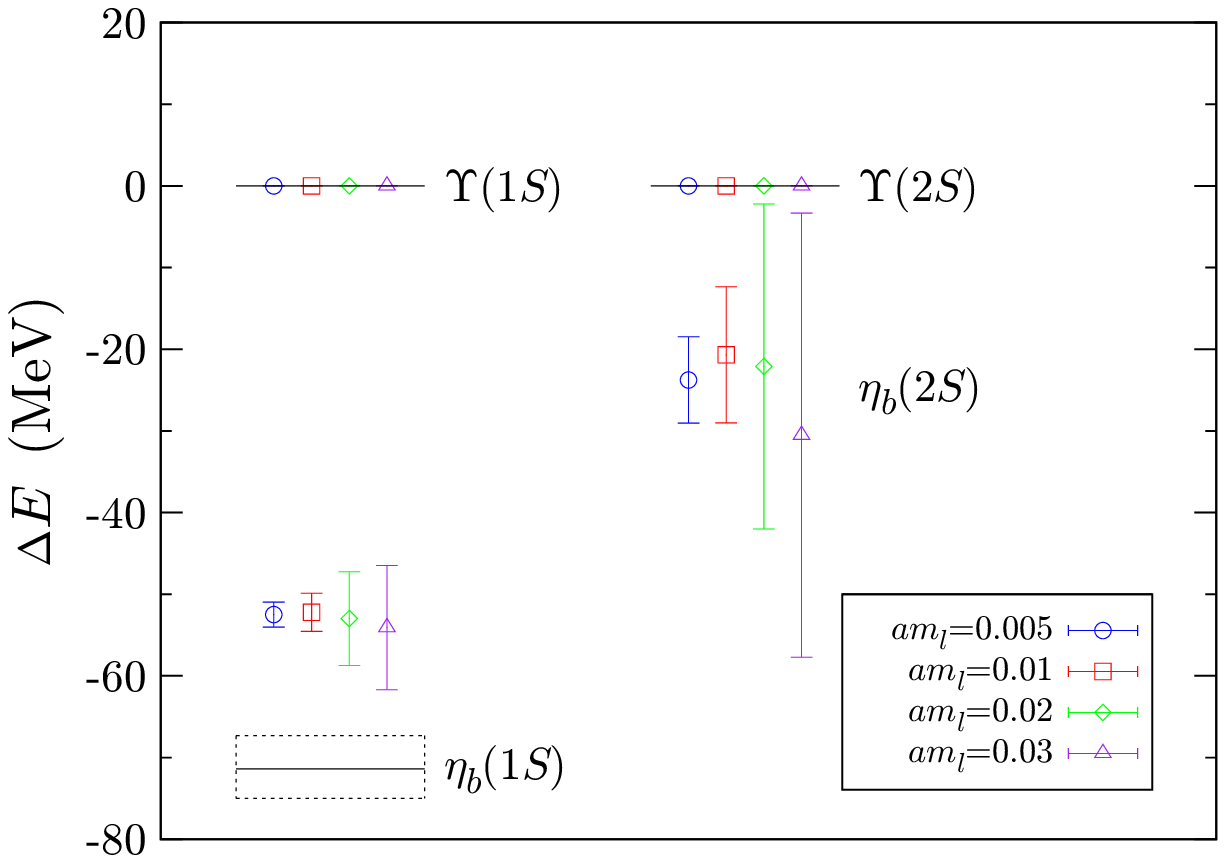}}
\caption{Bottomonium $S$-wave hyperfine splittings (energies relative to the
$\Upsilon(1S)$ and $\Upsilon(2S)$ states, respectively). Errors are statistical/fitting only.}
\label{fig:S_wave_hyperfine}
\end{minipage}
\hfill
\begin{minipage}{.48\linewidth}
\centerline{\includegraphics[width=\linewidth]{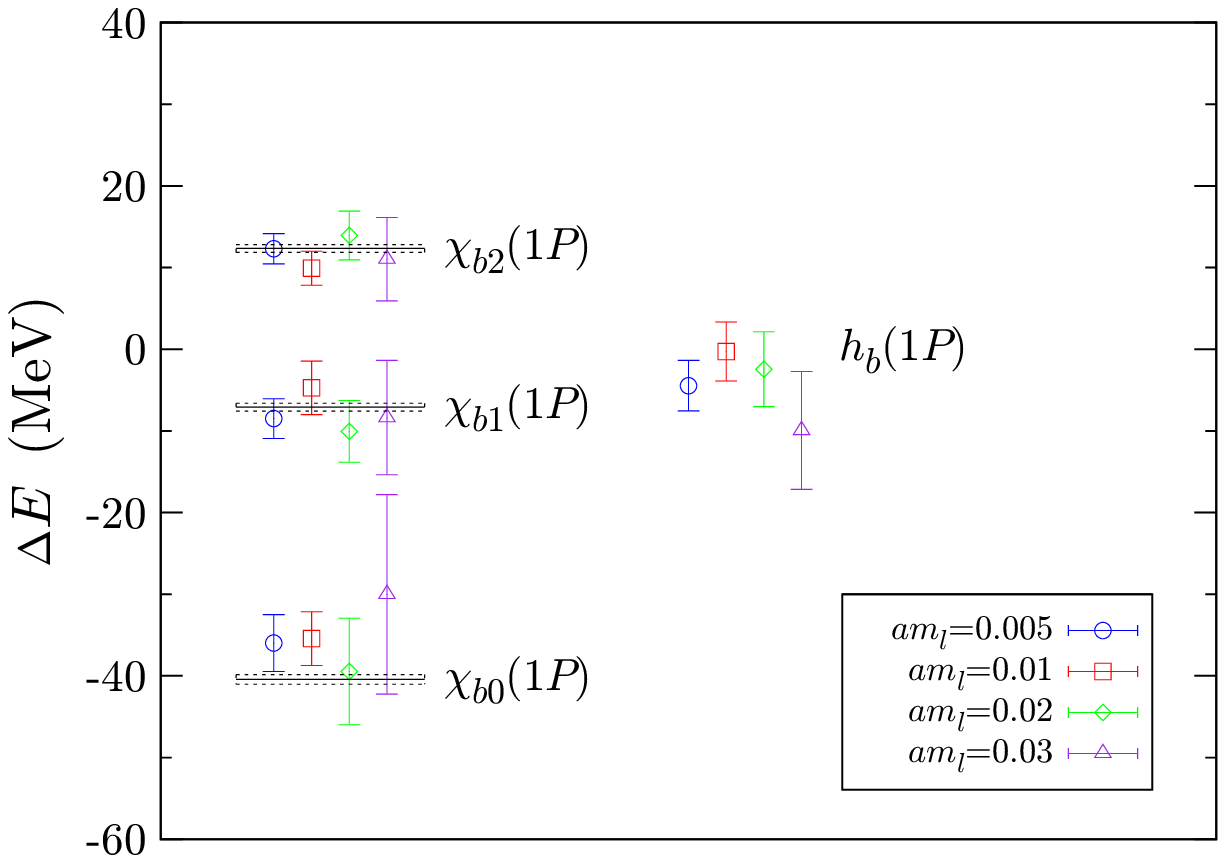}}
\caption{Bottomonium $P$-wave spin splittings (energies relative to the
spin-average of the $\chi_{b}(1P)$ states). Errors are statistical/fitting only.}
\label{fig:P_wave_spin_splittings}
\end{minipage}
\end{figure}

Results for the lattice spacings of the four different ensembles, computed after tuning the $b$ quark mass, are listed
in Table \ref{tab:lattice_spacing}. There, the first error given is statistical/fitting and the second is an estimate
of the systematic errors (relativistic, radiative and discretization) due to the NRQCD action.

Next, Fig.~\ref{fig:radial_orbital_vs_experiment} shows results for the radial and orbital energy splittings.
All masses have been determined by computing the energy difference to the $\Upsilon(1S)$ and using the experimental
$\Upsilon(1S)$ mass as an input. The lattice scales were taken from Table \ref{tab:lattice_spacing}.
Thus, the $2S$ and $1S$ masses are not predictions here and hence no error bars are shown for them. The
remaining energy splittings are in good agreement with the experimental results (lines). The sea quark mass dependence
is found to be weak, as expected for sufficiently light quarks.

Spin-dependent energy splittings were also computed and are shown in Figs.~\ref{fig:S_wave_hyperfine}
and \ref{fig:P_wave_spin_splittings}. Here, larger systematic errors are expected due to missing relativistic
and radiative corrections as well as discretization errors. The latter are most severe for the ($S$-wave) hyperfine
splitting, which is known to be sensitive to very short distances. The $P$-wave spin splittings shown in
Fig.~\ref{fig:P_wave_spin_splittings} are seen to be in relatively good agreement with experiment
within the statistical errors. The $1S$ hyperfine splitting was found to be $52.5\pm1.5({\rm stat})$ MeV
on the most chiral ensemble, which has to be compared to the experimental value of
$71.4^{+2.3}_{-3.1}({\rm stat})\pm2.7({\rm syst})$ MeV \cite{Aubert:2008vj}. In Ref.~\cite{Li:2008kb}, the
bottomonium spectrum was computed using a relativistic heavy-quark action on the same RBC/UKQCD gauge
configurations. There, the hyperfine splitting was found to be only $23.7\pm3.7({\rm stat})$ MeV, a much
larger deviation from experiment.

\section{Bottom mesons and baryons}
\label{sec:bottom_hadrons}

For the calculation of heavy-light meson and baryon masses, the set of $u/d$ and $s$ valence quark domain wall
propagators in use is an extension of the propagators that were computed and saved during the static-light
calculation in \cite{Detmold:2008ww}. So far, only propagators on the $am_l=0.005$, $am_s=0.04$ ensemble,
with valence quark masses equal to the sea quark masses have been included in the ongoing NRQCD spectrum
calculation. These quark masses correspond to pion and kaon masses of about $331$ and $576$ MeV, respectively.
Note that also the strange quark mass is too large; the physical point corresponds to $am_s\approx0.034$
\cite{Allton:2008pn}.

The domain wall propagators have APE smeared sources. For the heavy-quark, NRQCD propagators are computed with
both point and Gaussian smeared sources. Hadron correlation functions are then calculated for both point and
smeared sinks and projected to zero momentum. They are fitted simultaneously in a fully correlated
multi-exponential matrix fit, and errors are estimated using bootstrap. The results shown below are from
about 800 domain wall propagators. To increase statistics, correlators directed both forward and backward
in time are computed.

\begin{table}
\begin{minipage}{.5\linewidth}
\begin{center}
\begin{tabular}{lllllll}
\hline\hline
 \\[-2.5ex]
 Hadron                          & $J^P$                       & Operator \\
 \\[-2.5ex]
\hline
 \\[-2.5ex]
 $\Lambda_b$                     & $\sfrac12^+$                & $\epsilon_{abc}\:{(C\gamma_5)_{\beta\gamma}\:\:q^a_\beta\:\:q'^b_\gamma}\:\: {Q^c_\alpha}$ \\
 \\[-2.5ex]
 $\Sigma_b,$ $\Sigma_b^*$        & $\sfrac12^+,$ $\sfrac32^+$  & $\epsilon_{abc}\:{(C\gamma_j)_{\beta\gamma}\:\:q^a_\beta\:\:q'^b_\gamma}\:\: {Q^c_\alpha}$ \\
 \\[-2.5ex]
 $\Xi_b$                         & $\sfrac12^+$                & $\epsilon_{abc}\:{(C\gamma_5)_{\beta\gamma}\:\:q^a_\beta\:\:s^b_\gamma}\:\: {Q^c_\alpha}$  \\
 \\[-2.5ex]
 $\Xi_b',$ $\Xi_b^*$             & $\sfrac12^+,$ $\sfrac32^+$  & $\epsilon_{abc}\:{(C\gamma_j)_{\beta\gamma}\:\:q^a_\beta\:\:s^b_\gamma}\:\: {Q^c_\alpha}$  \\
 \\[-2.5ex]
 $\Omega_b,$ $\Omega_b^*$        & $\sfrac12^+,$ $\sfrac32^+$  & $\epsilon_{abc}\:{(C\gamma_j)_{\beta\gamma}\:\:s^a_\beta\:\:s^b_\gamma}\:\: {Q^c_\alpha}$  \\
 \\[-2.5ex]
 $\Xi_{bb},$ $\Xi_{bb}^*$        & $\sfrac12^+,$ $\sfrac32^+$  & $\epsilon_{abc}\:{(C\gamma_j)_{\beta\gamma}\:\:Q^a_\beta\:\:Q^b_\gamma}\:\: {q^c_\alpha}$  \\
 \\[-2.5ex]
 $\Omega_{bb},$ $\Omega_{bb}^*$  & $\sfrac12^+,$ $\sfrac32^+$  & $\epsilon_{abc}\:{(C\gamma_j)_{\beta\gamma}\:\:Q^a_\beta\:\:Q^b_\gamma}\:\: {s^c_\alpha}$  \\
 \\[-2.5ex]
 $\Omega_{bbb}$                  & $\sfrac32^+$                & $\epsilon_{abc}\:{(C\gamma_j)_{\beta\gamma}\:\:Q^a_\beta\:\:Q^b_\gamma}\:\: {Q^c_\alpha}$  \\
 \\[-2.5ex]
\hline\hline
\end{tabular}
\caption{\label{tab:baryon_ops}Operators for bottom baryons ($C=\gamma_4\gamma_2$, nonrelativistic gamma matrix basis, $m_u=m_d$).}
\end{center}
\end{minipage}
\hfill
\begin{minipage}{.45\linewidth}
\begin{center}
\begin{tabular}{ccc}
\hline\hline
  Splitting          &  $\displaystyle\begin{array}{c}
\Delta M\:\:\mathrm{(MeV)} \\ \mathrm{lattice} \end{array}$ & $\displaystyle\begin{array}{c}
\Delta M\:\:\mathrm{(MeV)} \\ \mathrm{experiment} \end{array}$ \\
\hline
  $B^* - B$                     &  $48(9)\phantom{0}$  &  $45.78(35)$  \\
  $B_s^* - B_s$                 &  $49(4)\phantom{0}$  &  $46.1(1.5)$  \\
  $\Sigma_b^* - \Sigma_b$       &  $25(25)$            &  $21.2(2.0)$  \\
  $\Xi_b^* - \Xi_b'$            &  $18(16)$            &  $-$ \\
  $\Omega_b^* - \Omega_b$       &  $19(10)$            &  $-$ \\
  $\Xi_{bb}^* - \Xi_{bb}$       &  $24(14)$            &  $-$ \\
  $\Omega_{bb}^* - \Omega_{bb}$ &  $38(9)\phantom{0}$  &  $-$ \\
\hline\hline
\end{tabular}
\caption{\label{tab:HL_spin_splittings}Heavy-light spin splittings in bottom mesons and baryons at
$am_l=0.005,\:\:\:am_s=0.04$ (preliminary; errors are statistical/fitting only)}
\end{center}
\end{minipage}
\end{table}

The structure of the baryon operators in use is shown in Table \ref{tab:baryon_ops}. The $b$ quark is denoted
by $Q$, which is a 4-component spinor with vanishing lower components (in the nonrelativistic gamma matrix basis),
since in NRQCD quarks and antiquarks are decoupled. The operators with Dirac matrix $\Gamma=C\gamma_j$ have an overlap
with both $J=\sfrac32$ and $J=\sfrac12$ states. At zero momentum, these contributions can be disentangled by
multiplying the correlator with the projectors $(\delta_{ij}-\sfrac13\gamma_i\gamma_j)$ and $\sfrac13\gamma_i\gamma_j$,
respectively.

\begin{figure}
\begin{minipage}{.48\linewidth}
\centerline{\includegraphics[width=\linewidth]{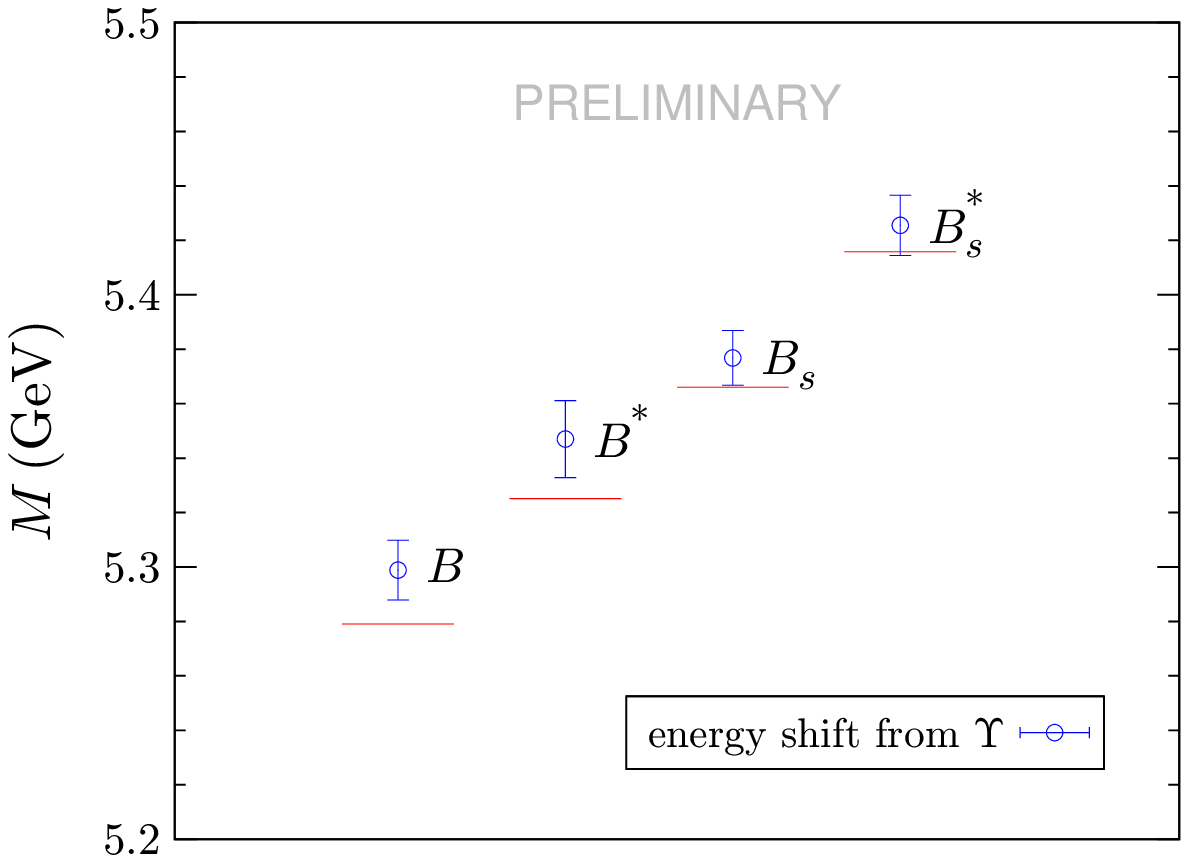}}
\caption{B meson masses at $am_l=0.005$, $am_s=0.04$. Errors are statistical/fitting only.}
\label{fig:B_mesons_vs_exp}
\end{minipage}
\hfill
\begin{minipage}{.48\linewidth}
\centerline{\includegraphics[width=\linewidth]{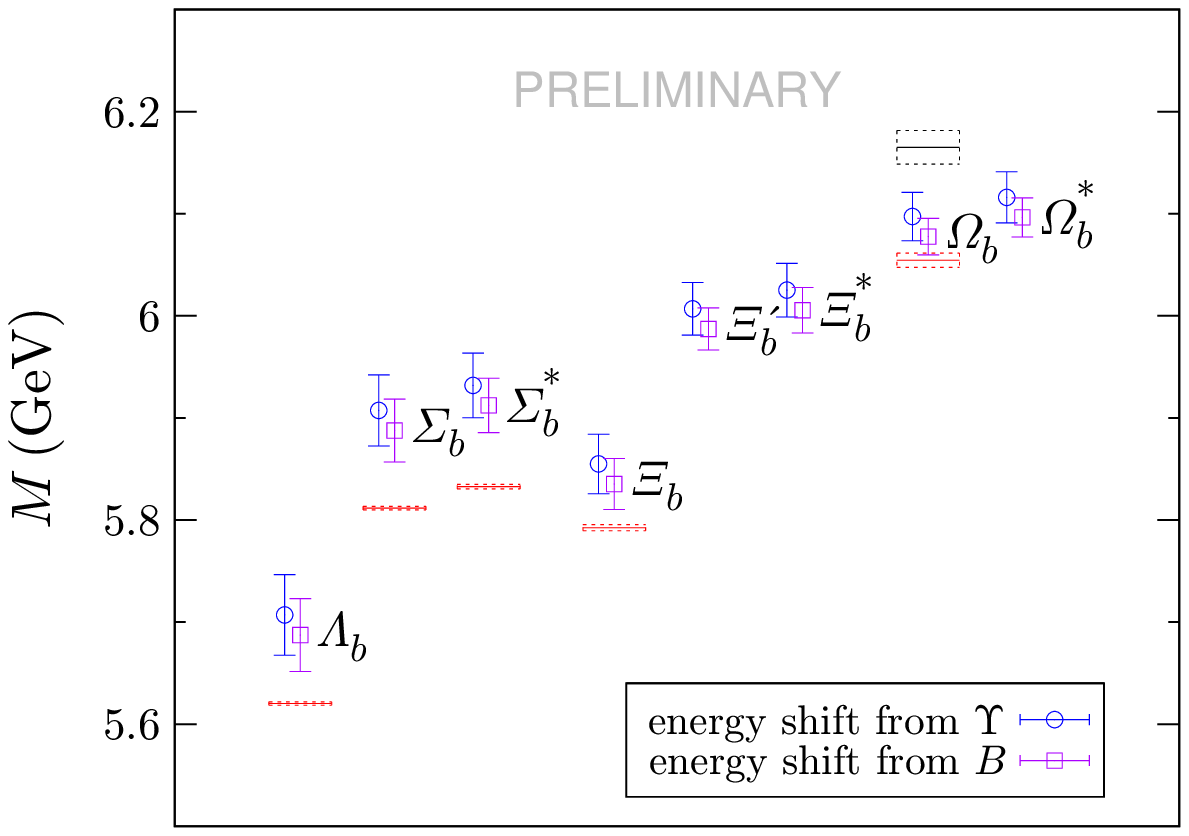}}
\caption{Singly bottom baryon masses at $am_l=0.005,\:\:\:am_s=0.04$. Errors are statistical/fitting only.}
\label{fig:singly_bottom_baryons_vs_exp}
\end{minipage}
\end{figure}

\begin{figure}
\begin{minipage}{.48\linewidth}
\centerline{\includegraphics[width=\linewidth]{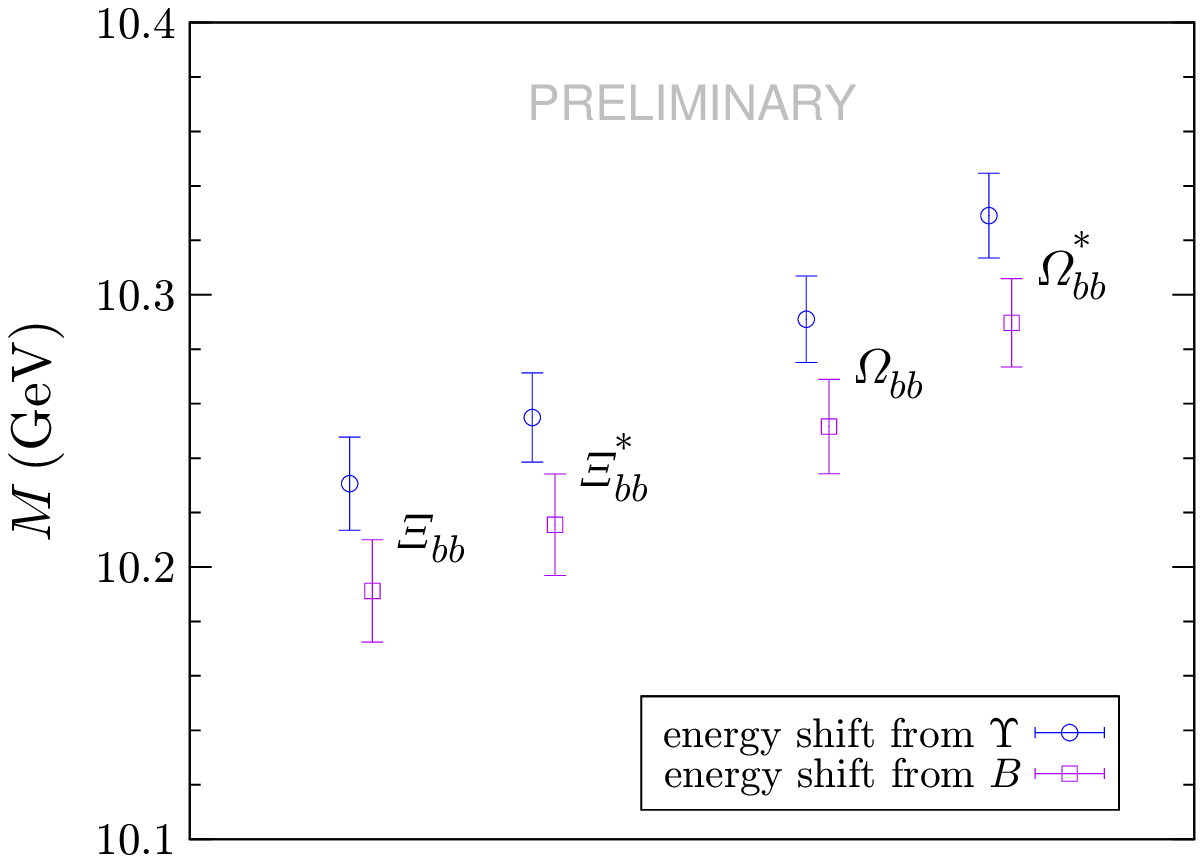}}
\caption{Doubly bottom baryon masses at $am_l=0.005,\:\:\:am_s=0.04$. Errors are statistical/fitting only.}
\label{fig:doubly_bottom_baryons}
\end{minipage}
\hfill
\begin{minipage}{.48\linewidth}
\centerline{\includegraphics[width=\linewidth]{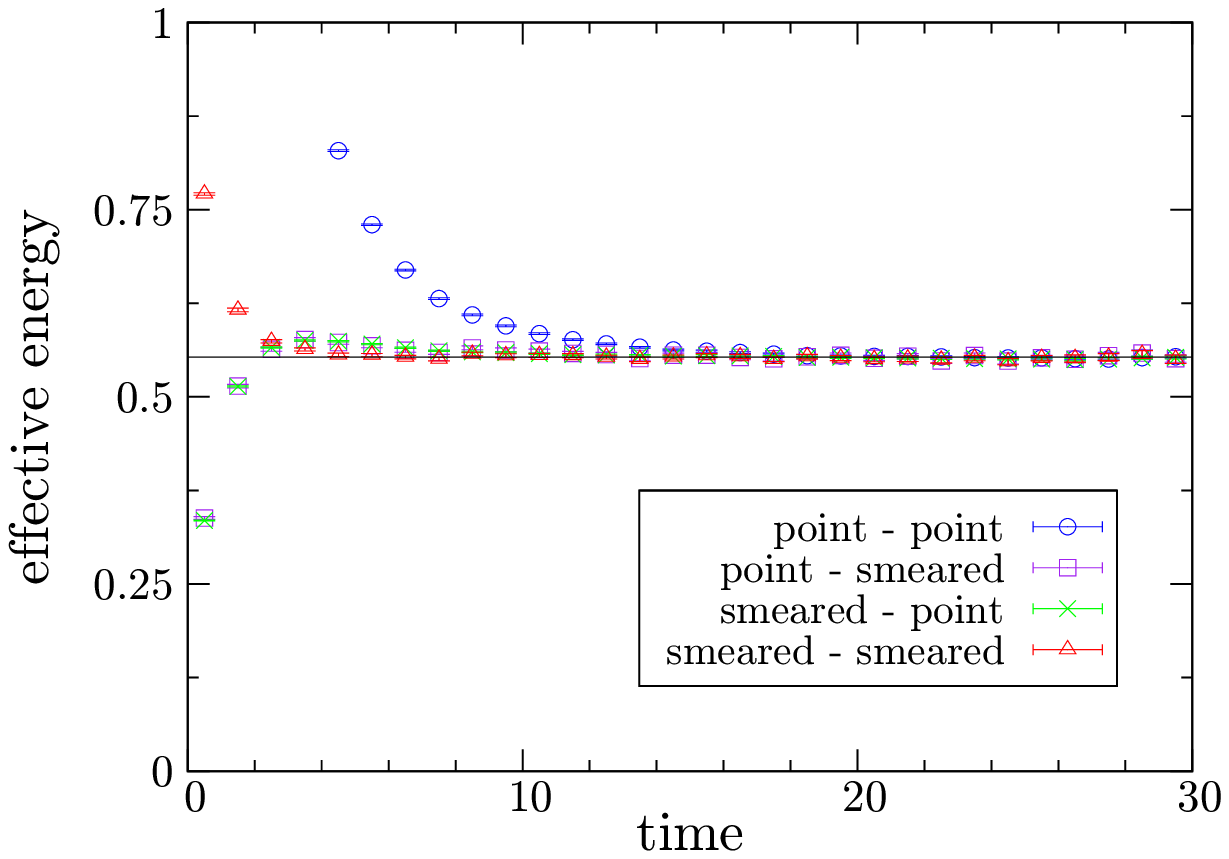}}
\caption{$\Omega_{bbb}$ matrix correlator, effective energy plot (lattice units).}
\label{fig:Omega_bbb_em}
\end{minipage}
\end{figure}

As mentioned before, energies obtained from fits to correlators are shifted due to the use of NRQCD. Energy splittings
are not affected. To compute the full hadron masses in a way that leads to only weak dependence on the bare $b$ quark
mass, the experimental value for the e.g. the $\Upsilon(1S)$ or the $B$ meson mass is used as an input parameter in
the following way:
\begin{eqnarray}
M&=&E_{\rm sim.}+\frac{{n_b}}{2}\left(M^{{\Upsilon}}_{\rm exp.}-E^{{\Upsilon}}_{\rm sim.} \right) \label{eq:Ups_method}\\
\mathrm{or}\hspace{2ex}M&=&E_{\rm sim.}+{n_b}\left(M^{{B}}_{\rm exp.}-E^{{B}}_{\rm sim.} \right) \label{eq:B_method},
\end{eqnarray}
where $n_b$ denotes the number of $b$ quarks in the hadron, $E_{\rm sim.}$ is the simulation energy and $M$ is the
full hadron mass to be calculated. Results for the $B$ meson masses, computed using (\ref{eq:Ups_method}), are shown
in Fig.~\ref{fig:B_mesons_vs_exp}, and the masses of singly- and doubly bottom baryons, for both methods
(\ref{eq:Ups_method}) and (\ref{eq:B_method}), are shown in Figs.~\ref{fig:singly_bottom_baryons_vs_exp} and
\ref{fig:doubly_bottom_baryons}. Where available, the experimental values are indicated \cite{Amsler:2008zzb};
for the $\Omega_b$, both the \Dzero (black) and CDF (red) results are shown \cite{Abazov:2008qm,Aaltonen:2009ny}.
Numerical results for various spin splittings are listed in Table \ref{tab:HL_spin_splittings}; these are found to
agree with experiment (where available) within the statistical errors. The hadron masses at the present values for
the light quark masses tend to be slightly above the experimental results. Definitive conclusions can only be
made after chiral extrapolation (and, eventually, after the inclusion of different lattice spacings and volumes).
Note that the $\Upsilon$ mass shows little dependence on the sea quark masses, while the $B$ has a light valence
quark. Thus, (\ref{eq:Ups_method}) and (\ref{eq:B_method}) lead to very different chiral behavior of $M$,
which likely explains the discrepancies between the two methods seen at the present quark masses (for doubly
bottom baryons, Fig.~\ref{fig:doubly_bottom_baryons}, the differences are enhanced since $n_b=2$).

The $\Omega_{bbb}$ baryon does not contain light valence quarks, and similarly to bottomonium, the dependence on
the light sea quarks masses is expected to be weak once these are light enough. Thus, Eq.~(\ref{eq:Ups_method})
is the better method for computing its absolute mass, and no chiral extrapolation is required. Also, since NRQCD
is computationally cheap, one can go to very high statistics with little cost. An effective-energy plot for an
$\Omega_{bbb}$ matrix correlator from about $10^5$ NRQCD propagators on the $am_l=0.005$, $am_s=0.04$ ensemble
is shown in Fig.~\ref{fig:Omega_bbb_em}. As can be seen, the signal is very good. The (unphysical) energy obtained
from the fit is $\:\:aE_{\Omega_{bbb}}=0.5527(12)$. Fitting an $\Upsilon$ correlator from the same propagators gives
$\:\:aE_{\Upsilon(1S)}=0.29786(20)$. Using the bootstrap method to properly take into account correlations,
Eq.~(\ref{eq:Ups_method}) then leads to
\begin{equation}
 M_{\Omega_{bbb}}=14.3748(33)\:\:\mathrm{GeV}
\end{equation}
where the error is statistical only and includes the uncertainty in the lattice spacing (the latter was taken
from Table \ref{tab:lattice_spacing}). The $\Omega_{bbb}$ mass has been estimated using various continuum methods,
see \cite{Hasenfratz:1980ka, Jia:2006gw, Bernotas:2008bu, Martynenko:2007je, Zhang:2009re}, and the production
of the $\Omega_{bbb}$ at hadron colliders has been studied in \cite{GomshiNobary:2004mq, GomshiNobary:2005ur}.

\section{Outlook}

The heavy-light calculations will be extended to include the other light quark masses, and chiral
extrapolations will be performed. All calculations presented here are only for one lattice spacing,
but the finer $V=32^3\times64$ RBC/UKQCD gauge configurations will be included once they become
available. This should allow more reliable estimates of discretization errors.

With NRQCD, high statistical accuracy can be achieved for the $\Omega_{bbb}$ baryon, similarly to bottomonium.
It should therefore be possible to study excited states also for the $\Omega_{bbb}$.

\vspace{2ex}

\noindent \textbf{Acknowledgments:} Computations were performed at NCSA, NERSC, and Cambridge HPCS.

\vspace{-0.5ex}

\providecommand{\href}[2]{#2}

\end{document}